\documentclass[two column, prA]{revtex4}%
\usepackage{amsmath}
\usepackage{graphicx}
\usepackage{amsfonts}
\usepackage{amssymb}
\usepackage{color}%
\providecommand{\U}[1]{\protect\rule{.1in}{.1in}}
\ifx\pdfoutput\relax\let\pdfoutput=\undefined\fi
\newcount\msipdfoutput
\ifx\pdfoutput\undefined\else
\ifcase\pdfoutput\else
\ifx\paperwidth\undefined\else
\ifdim\paperheight=0pt\relax\else\pdfpageheight\paperheight\fi
\ifdim\paperwidth=0pt\relax\else\pdfpagewidth\paperwidth\fi
\fi\fi\fi
\begin{document}
\title{Maxwell meets Reeh-Schlieder: the quantum mechanics of neutral bosons}
\author{Margaret Hawton}
\affiliation{Department of Physics, Lakehead University, Thunder Bay, ON, Canada, P7B 5E1}
\email{margaret.hawton@lakeheadu.ca}
\author{Vincent Debierre}
\affiliation{Max Planck Institute for Nuclear Physics, Saupfercheckweg 1, 69117,
Heidelberg, Germany}
\email{debierrev@mpi-hd.mpg.de}

\begin{abstract}
We find that biorthogonal quantum mechanics with a scalar product that counts
both absorbed and emitted particles leads to covariant position operators with
localized eigenvectors. In this manifestly covariant formulation the
probability for a transition from a one-photon state to a position eigenvector
is the first order Glauber correlation function, bridging the gap between
photon counting and the sensitivity of light detectors to electromagnetic
energy density. The position eigenvalues are identified as the spatial
parameters in the canonical quantum field operators and the position basis
describes an array of localized devices that instantaneously absorb and
re-emit bosons.

\end{abstract}
\maketitle

\section{Introduction}

In nonrelativistic quantum mechanics (QM) the wave function is the projection
of a particle's state vector onto a basis of position eigenvectors and its
absolute square is the positive definite probability density. However, many
experimental tests of QM are performed on photons and there is currently no
well-defined relativistic QM \cite{Nikolic,Pauli,Redhead,Malament,Halvorson}
of photons or the neutral Klein-Gordon (KG) bosons often considered in their
place for simplicity \cite{BohmHiley,BirrellDavies}. Recently it was claimed
that the photon wave function \cite{Lundeen} and Bohmian photon trajectories
\cite{Steinberg} were observed using weak measurements. This interpretation,
justified by the analogy between the paraxial and Schr\"{o}dinger equations,
is disputed and an alternative interpretation based on the electromagnetic
field has been presented \cite{Hiley,Holland93,BohmHiley}. In quantum optics
most theorists deny the existence of number density and instead base
calculations on energy density \cite{ScullyZubairy,Sipe,BB96,SmithRaymer},
although QM based on number density has been proposed \cite{Hawton07}. The
photon wave function and its application to emission by an atom and Bohmian
trajectories is reviewed in \cite{Keller,Debierre}. Two sources of nonlocality
have contributed to the perception that there is no relativistic QM or number
density: Wave functions are assumed to be of positive frequency while
Hegerfeldt's theorem \cite{Hegerfeldt} tells us that this restriction leads to
instantaneous spreading, and the Newton-Wigner (NW) position eigenvectors
\cite{NewtonWigner} are localized in the sense that they are orthogonal but
their relationship to the physical fields and to current sources is nonlocal
in configuration space. We will argue here that both of these sources of
nonlocality are nonphysical: In biorthogonal QM \cite{Brody} the nonlocal
transformation to the NW basis is not required \cite{HD} and a scalar product
exists \cite{BabaeiMostafazadeh} that does not require separation of the
fields into their nonlocal \cite{Halvorson} positive and negative frequency
parts \cite{WestraKoksma}. As a consequence real fields are allowed and the
paradoxical observer dependence of particle density on acceleration
\cite{Unruh,BirrellDavies} can be avoided. In the manifestly covariant
formalism derived here we identify the position eigenvalues of relativistic QM
with the spatial parameters in the canonical quantum field theory (QFT)
operators and the localized states as derivatives of Green functions that
describe an array of emitting and absorbing devices localized in spacetime.
This unifies the physical interpretation of the position coordinate in
classical electromagnetism, relativistic QM and QFT.

The conventional scalar product \cite{BirrellDavies} is a difference of
particle and antiparticle terms so it is indefinite unless the field is
limited to positive frequencies. The scalar product derived in
\cite{Mostafazadeh1,BabaeiMostafazadeh} is positive definite for both positive
and negative frequency fields. If this scalar product is used, inclusion of
negative frequency states becomes mathematically straightforward. First
quantized fields describing neutral KG particles and photons are real so
inclusion of negative frequency fields is not only reasonable, it is
essential. For photons the first quantized theory derived in
\cite{BabaeiMostafazadeh} restricted to real fields is essentially classical
electrodynamics with a rule for calculating number density. However, QFT is
required for description of multiphoton states and entanglement. It is a
consequence of the Reeh-Schlieder (RS) theorem \cite{RS} of algebraic QFT
(AQFT) \cite{HaagKastler} that there are no local creation or annihilation
operators and the vacuum is entangled across spacelike separated regions
\cite{Halvorson,Redhead,Malament,Summers}. Any number operator that counts a
particle by annihilating it with the positive frequency part of a field
operator and then recreating it is nonlocal. Localization is a finite region
requires summation over positive and negative frequencies \cite{Leon}. A
particle's energy must be bounded from below to prevent it from acting as an
infinite energy source \cite{Malament} so it is essential to make a
distinction between energy and frequency. In AQFT, also called local QFT,
causality is enforced by the microcausality assumption that field operators
defined across space-like separated regions commute \cite{Redhead}. In
Sections 3 and 4 both positive and negative frequency states will be defined
and in Section 6 their interpretation in terms of absorbed and emitted
positive energy particles will be elaborated on. We find no conflict of our
localized bases with AQFT; rather the RS theorem and microcausality support
our proposal that, in a covariant formulation, both absorbed and emitted
neutral bosons should be counted.

The wave functions derived in \cite{Mostafazadeh1,BabaeiMostafazadeh} are
projections of the field onto NW position eigenvectors \cite{NewtonWigner}. NW
found a position operator for KG particles, but they concluded that the only
photon position operator is the Pryce operator whose vector components do not
commute \cite{Pryce}, making the simultaneous determination of photon position
in all three directions of space impossible. They had assumed spherically
symmetrical position eigenstates for photons, while photon position
eigenvectors have an axis of symmetry like twisted light \cite{Twisted}. The
photon Poincar\'{e} group is discussed in \cite{HawtonPosOp,HawtonBaylis}.
Following the NW method with omission of the spherical symmetry axiom, a
photon position operator with commuting components and cylindrically
symmetrical eigenvectors can be constructed \cite{HawtonPosOp}. Since spin and
orbital angular momentum are not separately observable \cite{CohenQED1}, its
eigenvectors have only definite total angular momentum along some fixed but
arbitrary axis \cite{HawtonBaylis}. A generalization of this cylindrically
symmetric NW position operator was derived independently in
\cite{BabaeiMostafazadeh}. Here we retain this symmetry but omit the NW
similarity transformation that leads to nonlocality in configuration space.

The similarity transformation to the NW basis preserves scalar products but
its nonlocal relationship to the physical fields has been interpreted as a
nonlocal relationship between number density and energy density \cite{BB96}.
The nontrivial metric factor in the NW basis is not physically observable
\cite{Brody2}, and this suggests that nonlocality of the NW position
eigenvectors is also not physically observable. Here we will work in the
formalism of biorthogonal QM \cite{Brody} that does not require transformation
to the NW basis. Biorthogonal QM in a finite dimensional Hilbert space is
summarized here as follows: The eigenvectors of a quasi-Hermitian \cite{quasi}
operator $\hat{O}$ and its adjoint $\hat{O}^{\dagger}$ are not orthogonal, as
is the case for conventional Hermitian operators, but biorthogonal. This means
that, given the eigenvector equations%
\begin{align}
\hat{O}|\omega_{i}\rangle &  =\omega_{i}|\omega_{i}\rangle,\\
\hat{O}^{\dagger}|\tilde{\omega}_{j}\rangle &  =\omega_{j}|\tilde{\omega}%
_{j}\rangle
\end{align}
we have $\langle\tilde{\omega}_{j}|\omega_{i}\rangle=\delta_{ji}\langle
\tilde{\omega}_{i}|\omega_{i}\rangle$ and the completeness relation
$\widehat{1}=\sum_{i}\left\vert \omega_{i}\right\rangle \left\langle
\widetilde{\omega}_{i}\right\vert /\langle\tilde{\omega}_{i}|\omega_{i}%
\rangle.$ An arbitrary state $\left\vert \psi\right\rangle $ has an associated
state $\left\vert \widetilde{\psi}\right\rangle $. If an arbitrary state
vector is expanded as $\left\vert \psi\right\rangle =\sum_{i}c_{i}\left\vert
\omega_{i}\right\rangle $ in the Hilbert space $\mathcal{H}$ then in
biorthogonal QM its associated state is $\left\vert \widetilde{\psi
}\right\rangle =\sum_{i}c_{i}\left\vert \widetilde{\omega}_{i}\right\rangle
\in\mathcal{H}^{\ast}$ where $c_{i}=\left\langle \widetilde{\omega}_{i}%
|\psi\right\rangle =\left\langle \omega_{i}|\widetilde{\psi}\right\rangle .$
Using these expansions it is straightforward to verify that $\left\langle
\widetilde{\psi}_{1}|\psi_{2}\right\rangle =\left\langle \psi_{1}%
|\widetilde{\psi}_{2}\right\rangle .$ The probability for a transition from a
quantum state $\left\vert \psi\right\rangle $ to an eigenvector $\left\vert
\tilde{\omega}_{i}\right\rangle ${\ of }$\hat{O}^{\dagger}$ is
\begin{equation}
p_{i}=\frac{\left\vert \left\langle \widetilde{\omega}_{i}|\psi\right\rangle
\right\vert ^{2}}{\left\langle \widetilde{\psi}|\psi\right\rangle \left\langle
\widetilde{\omega}_{i}|\omega_{i}\right\rangle }. \label{p}%
\end{equation}
A generic operator can be written in the form
\begin{equation}
\widehat{F}=\sum_{i,j}f_{ij}\left\vert \omega_{i}\right\rangle \left\langle
\widetilde{\omega}_{j}\right\vert \label{F}%
\end{equation}
where $f_{ij}=\left\langle \widetilde{\omega}_{i}\left\vert \widehat{F}%
\right\vert \omega_{j}\right\rangle $ can be viewed as a matrix \cite{Brody}.
An equivalent bottom up approach is to start with a set of linearly
independent not necessarily orthogonal vectors and obtain a biorthogonal basis
and operators describing observables \cite{Brody2}. In Section 3 we will apply
this formalism to the biorthogonal position eigenvectors $\left\vert
\phi\left(  x\right)  \right\rangle =\widehat{\phi}\left(  x\right)
|0\rangle$ and $\left\vert \widetilde{\phi}\left(  x\right)  \right\rangle
=\left\vert \pi\left(  x\right)  \right\rangle \propto\widehat{\pi}\left(
x\right)  |0\rangle$ where $x^{\mu}=\left(  ct,\mathbf{x}\right)  ,$
$\widehat{\phi}\left(  x\right)  $ is a field operator, $\widehat{\pi}\left(
x\right)  $ is its conjugate momentum operator, and $|0\rangle$ is the global
vacuum state. Extension of biorthogonal QM to this infinite-dimensional
Hilbert space is not rigorous; for example completeness could fail as
discussed in \cite{Brody}.

{The rest of this paper is organized as follows: In Section 2 }KG wave
mechanics,{\ with the field rescaled here to facilitate application to
particles with zero mass, is reviewed. In Section~3} the covariant position
operator and positive definite probability density are derived. {In Section~4
the KG position observable discussed in Sections 2 and 3 is extended to
photons. In Section~5 the wave function of the photon emitted by an atom is
discussed, in Section 6 inclusion of negative frequency states, causality and
localized states are examined, and in Section~7 we conclude.}

\section{Klein-Gordon wave mechanics}

We will start with a review of the KG position observable problem. The KG
equation%
\begin{equation}
\partial_{\mu}\partial^{\mu}\phi\left(  x\right)  +\frac{m^{2}c^{2}}{\hbar
^{2}}\phi\left(  x\right)  =0 \label{KGeq}%
\end{equation}
describes charged and neutral particles with zero spin (pions). Here covariant
notation and the mostly minus convention are used in which $x^{\mu}=x=\left(
ct,\mathbf{x}\right)  ,$ $\partial_{\mu}=\left(  \partial_{ct},\mathbf{\nabla
}\right)  ,$ $m$ is the mass of the KG particle, $c$ is the speed of light,
$2\pi\hbar$ is Planck's constant and $f_{1}\overleftrightarrow{\partial}_{\mu
}f_{2}\equiv f_{1}\left(  \partial_{\mu}f_{2}\right)  -\left(  \partial_{\mu
}f_{1}\right)  f_{2}.$ The function $\phi\left(  x\right)  $ is any scalar
field that satisfies the KG equation (\ref{KGeq}).\textbf{\ }The four-density
\begin{equation}
J_{KG}^{\mu}\left(  x\right)  =\mathrm{i}g\phi\left(  x\right)  ^{\ast
}\overleftrightarrow{\partial}^{\mu}\phi\left(  x\right)  , \label{KG}%
\end{equation}
satisfies a continuity equation. Plane wave normal mode solutions to
(\ref{KGeq}) proportional to $\exp\left(  -\mathrm{i}\omega t\right)  $ are
referred to as positive frequency solutions, while those proportional to
$\exp\left(  \mathrm{i}\omega t\right)  $ are negative frequency. Completeness
requires that both positive and negative frequency modes be included. Their
contributions to $J_{KG}^{0}\left(  x\right)  $ are of opposite sign, so
$J_{KG}^{0}\left(  x\right)  $ is interpreted as charge density and the
quantity $g$ in (\ref{KG}) is set equal to $qc/\hbar$ for particles of charge
$q$.

If only particles, {as opposed to both particles and antiparticles}, are to be
considered, then the KG field can be restricted to positive frequencies and
the scalar product \cite{BirrellDavies}%
\begin{equation}
\left(  \phi_{1},\phi_{2}\right)  _{KG}=\frac{\mathrm{i}}{\hbar}\int%
_{t}\mathrm{d}\mathbf{x}\,\phi_{1}\left(  x\right)  ^{\ast}%
\overleftrightarrow{\partial}_{t}\phi_{2}\left(  x\right)  \label{KGsp}%
\end{equation}
is positive definite. Here $t$ denotes a spacelike hyperplane of simultaneity
at instant $t.$ The integrand of (\ref{KGsp}) looks like a particle density
but this is misleading since $J_{KG}^{0}\left(  x\right)  $ can still be
negative if components with two or more different frequencies are added
\cite{Holland,Colin,Nikolic}. Restriction to positive frequencies is
especially problematic in the case of neutral particles that should be
described by real fields.

The problem of a probability interpretation for KG particles has a long
history. Lack of a probability interpretation led Dirac to derive his
celebrated equation for spin half particles, but this does not solve the
problem for KG fields. In a seminal paper intended to clarify the confusion
about relativistic wave mechanics, Feshbach and Villars reviewed the two
component formalism that separates the wave function into its particle and
antiparticle parts for charged or for neutral particles \cite{FeshbachVillars}%
. Their treatment of neutral particles was based on energy density. Since this
work various strategies have been employed to derive a positive definite
probability density. The four-current density can be redefined so that its
zeroth component is positive definite \cite{GGP}, but this construction has no
apparent physical basis and it fails if $m=0$ \cite{HenningWolf}. It has been
proposed that for charged pions only positive definite eigenstates of the
Hamiltonian are physical \cite{SUL}. A new $J^{\mu}\left(  x\right)  $ was
derived that does not require separation of the field into positive and
negative frequency parts \cite{Mostafazadeh1,Mostafazadeh2} so it can be
applied to the real fields that describe neutral pions. If $\phi\left(
x\right)  $ is restricted to positive frequencies it reduces to (\ref{KG}) so
this $J^{\mu}\left(  x\right)  $ that describes an arbitrary linear
combination of positive and negative frequency fields, including real fields,
will be used here.

Working in the two component formalism with a pseudo-Hermitian Hamiltonian
Mostafazadeh \cite{Mostafazadeh1,Mostafazadeh2} defined the positive-definite
Hermitian operator
\begin{equation}
\widehat{D}\equiv-\nabla^{2}+m^{2}c^{2}/\hbar^{2} \label{D}%
\end{equation}
in terms of which the KG equation is $\left(  \widehat{D}+\partial_{ct}%
^{2}\right)  \psi=0.$ He derived the conjugate field
\begin{equation}
\phi_{c}\left(  x\right)  =\mathrm{i}\widehat{D}^{-1/2}\partial_{ct}%
\phi\left(  x\right)  \label{psi_c}%
\end{equation}
such that if $\phi=\phi^{+}+\phi^{-}$ then $\phi_{c}=\phi^{+}-\phi^{-}.$ This
implies that $\phi_{c}$ is a scalar. It is then straightforward to verify that
$\phi_{c}$ satisfies the KG equation and that%
\begin{equation}
J^{\mu}\left(  x\right)  =\frac{\mathrm{i}}{\hbar}\phi\left(  x\right)
^{\ast}\overleftrightarrow{\partial}^{\mu}\phi_{c}\left(  x\right)  \label{M}%
\end{equation}
satisfies the continuity equation $\partial_{\mu}J^{\mu}=0$. Up to a constant
that just scales $J^{\mu},$ (\ref{M}) is the expression derived in
\cite{Mostafazadeh2}. Like (\ref{KG}) $J^{\mu}\left(  x\right)  $ is
manifestly covariant. It was proved in \cite{Mostafazadeh1,Mostafazadeh2} that
the scalar product
\begin{equation}
\left(  \phi_{1},\phi_{2}\right)  =\frac{\mathrm{i}}{\hbar}\int_{t}%
\mathrm{d}\mathbf{x}\,\phi_{1}\left(  x\right)  ^{\ast}%
\overleftrightarrow{\partial}_{t}\phi_{2c}\left(  x\right)  \label{Msp}%
\end{equation}
is positive definite, time independent and can be written in covariant form
as
\begin{equation}
\left(  \phi_{1},\phi_{2}\right)  =\frac{\mathrm{i}}{\hbar}\int_{n}%
{\mathrm{d}}\sigma n_{\mu}\phi_{1}\left(  x\right)  ^{\ast}%
\overleftrightarrow{\partial}^{\mu}\phi_{2c}\left(  x\right)
\label{invariant_sp}%
\end{equation}
where $n$ is an arbitrary spacelike hyperplane with normal $n^{\mu}$, in other
words, it is a Cauchy surface. The infinitesimal volume elements
{${\mathrm{d}}\sigma\equiv\mathrm{d}x$} are invariant. When restricted to
positive frequencies (\ref{M}) reduces to (\ref{KG}) and (\ref{Msp}) reduces
to (\ref{KGsp}). Using (\ref{KGeq}), (\ref{psi_c}) and $\phi_{c}=\phi^{+}%
-\phi^{-}$, the scalar product (\ref{Msp}) can be written as
\begin{equation}
\left(  \phi_{1},\phi_{2}\right)  =\frac{2c}{\hbar}\sum_{\epsilon=\pm
}\left\langle \phi_{1}^{\epsilon}|\widehat{D}^{1/2}\phi_{2}^{\epsilon
}\right\rangle \label{sp}%
\end{equation}
where%
\begin{equation}
\left\langle \chi_{1}|\chi_{2}\right\rangle =\int\mathrm{d}\mathbf{x}%
\,\chi_{1}^{\ast}\left(  \mathbf{x}\right)  \chi_{2}\left(  \mathbf{x}\right)
. \label{nonrel_sp}%
\end{equation}

The non-relativistic Hilbert space is the vector space of square integrable
continuous functions with the scalar product (\ref{nonrel_sp}). In the
relativistic Hilbert space the scalar product used here is (\ref{sp}). These
scalar products can be evaluated in configuration space or in $\mathbf{k}%
$-space.\textbf{\ }The covariant Fourier transform is%
\begin{equation}
\phi^{\epsilon}\left(  x\right)  =\int\frac{\mathrm{d}\mathbf{k}\,}{\left(
2\pi\right)  ^{3}2\omega_{\mathbf{k}}}\pi^{\epsilon}\left(  \mathbf{k}\right)
\mathrm{e}^{-\mathrm{i}\epsilon\left(  \omega_{\mathbf{k}}t-\mathbf{k\cdot
x}\right)  }. \label{FT}%
\end{equation}
Since $\phi^{\epsilon}\left(  x\right)  $ is a scalar and $\mathrm{d}%
\mathbf{k}\,/\left[  \left(  2\pi\right)  ^{3}2\omega_{\mathbf{k}}\right]  $
is invariant, $\pi^{\epsilon}\left(  \mathbf{k}\right)  \in\mathcal{H}^{\ast}$
is a scalar, analogous to the transformation properties of photons
\cite{HawtonLorentzInv08}. For $\omega_{\mathbf{k}}=\sqrt{\mathbf{k}^{2}%
c^{2}+m^{2}c^{4}/\hbar^{2}}$ the function $\phi^{\epsilon}\left(  x\right)  $
satisfies the KG equation and (\ref{sp}) can be written as
\begin{equation}
\left(  \phi_{1},\phi_{2}\right)  =\frac{1}{\hbar}\sum_{\epsilon=\pm}\int%
\frac{\mathrm{d}\mathbf{k}\,}{\left(  2\pi\right)  ^{3}2\omega_{\mathbf{k}}%
}\pi_{1}^{\epsilon\ast}\left(  \mathbf{k}\right)  \pi_{2}^{\epsilon}\left(
\mathbf{k}\right)  . \label{k_sp}%
\end{equation}
In the biorthogonal formalism the bases $\left\{  \phi_{\mathbf{x}_{j}%
}^{\epsilon}\left(  \mathbf{k}\right)  \right\}  =\left\{  \pi_{\mathbf{x}%
_{j}}^{\epsilon}\left(  \mathbf{k}\right)  /\omega_{\mathbf{k}}\right\}
\in\mathcal{H}$ and $\left\{  \pi_{\mathbf{x}_{j}}^{\epsilon}\left(
\mathbf{k}\right)  \right\}  \in\mathcal{H}^{\ast}$ are biorthogonal and
complete and the Hermitian adjoint of an operator is its complex conjugate
transpose \cite{Brody}. Since the scalar product (\ref{sp}) is positive
definite, standard QM can be recovered if a nontrivial metric
operator\textbf{\ }$\left\langle .|\Theta.\right\rangle $\textbf{\ }is
introduced \cite{GeyerScholtz}. In this metric formulation the basis is
$\left\{  \pi_{\mathbf{x}_{j}}^{\epsilon}\left(  \mathbf{k}\right)  \right\}
$ and operators are Hermitian. With the metric\textbf{\ }$\left\langle
.|.\right\rangle $\textbf{\ }operators representing observables can be
non-Hermitian with biorthogonal eigenvectors. The norm and orthogonality of
the elements of the Hilbert space\ and the concept of Hermiticity are
determined by the definition of scalar product. Newton and Wigner defined the
KG fields $\pi_{\mathbf{x}_{j}}^{\epsilon}\left(  \mathbf{k}\right)  \propto
e^{-\mathrm{i}\mathbf{k\cdot x}_{j}}\omega_{\mathbf{k}}^{1/2}$ that satisfy
$\left(  \phi_{1},\phi_{2}\right)  \propto\delta\left(  \mathbf{x}%
_{1}-\mathbf{x}_{2}\right)  \ $and are eigenvectors of the position operator
$\mathrm{i}\nabla_{\mathbf{k}}-\frac{\mathrm{i}c^{2}}{2\omega_{\mathbf{k}}%
^{2}}\mathbf{k}$ \cite{NewtonWigner}.\textbf{\ }Hermiticity of the NW position
operator with eigenvectors of this form is discussed by Pike and Sarkar
\cite{PikeSarkar}. The NW position operator has played a central role in the
discussion of relativistic particle position since its publication in 1949
\cite{NewtonWigner}, but this operator is not covariant and its eigenvectors
are not localized. We will show in the next section that the formalism of
biorthogonal QM leads to a covariant position operator and in Section 6 that
the canonical commutation relations are consistent with (\ref{sp}). If limited
to the calculation of scalar products and expectation values the biorthogonal
formalism is completely equivalent to the Hermitian formalism with a
nontrivial metric, so the role biorthogonal QM is to avoid falling into the
trap of treating NW nonlocality as a physically observable effect.

\section{KG position eigenvectors}

In this Section the second quantized formulation will be discussed first to
motivate the definition of the biorthogonal basis. Positive and negative
frequency basis states are then defined that provide a configuration space
basis for first or second quantized states. A completeness relation, position
operator and particle number density are derived within the framework of
biorthogonal QM. The relationship of this covariant position operator to the
Hermitian NW position operator is examined.

In QFT particles are created at a point in spacetime by a field operator or
its canonical conjugate. The interaction picture (IP) scalar field operators
$\widehat{\phi}\left(  x\right)  $ and $\widehat{\pi}\left(  x\right)
=\partial_{t}\widehat{\phi}\left(  x\right)  $ will be written as%
\begin{align}
\widehat{\phi}\left(  x\right)   &  =\sqrt{\hbar}\int\frac{\mathrm{d}%
\mathbf{k}}{\left(  2\pi\right)  ^{3}2\omega_{\mathbf{k}}}\mathrm{e}%
^{\mathrm{i}\left(  \omega_{\mathbf{k}}t-\mathbf{k}\cdot\mathbf{x}\right)
}\widehat{a}^{\dagger}\left(  \mathbf{k}\right)  +\mathrm{H.c.},\label{phi}\\
\widehat{\pi}\left(  x\right)   &  =\mathrm{i}\sqrt{\hbar}\int\frac
{\mathrm{d}\mathbf{k}}{\left(  2\pi\right)  ^{3}2}\mathrm{e}^{\mathrm{i}%
\left(  \omega_{\mathbf{k}}t-\mathbf{k}\cdot\mathbf{x}\right)  }%
\widehat{a}^{\dagger}\left(  \mathbf{k}\right)  +\mathrm{H.c.}%
\end{align}
where $\mathrm{H.c.}$ is the Hermitian conjugate\textbf{\ }and the
$\mathbf{k}$-space covariant commutation relations are \cite{ItzyksonZuber}
\begin{equation}
\left[  \widehat{a}\left(  \mathbf{k}\right)  ,\widehat{a}^{\dagger}\left(
\mathbf{q}\right)  \right]  =\left(  2\pi\right)  ^{3}2\omega_{\mathbf{k}%
}\,\delta\left(  \mathbf{k}-\mathbf{q}\right)  . \label{k_communation}%
\end{equation}
On the $t$ hyperplane the field operators satisfy the commutation relations
\begin{equation}
\left[  \widehat{\phi}\left(  x\right)  ,\widehat{\pi}\left(  y\right)
\right]  =\mathrm{i}\hbar\delta\left(  \mathbf{x}-\mathbf{y}\right)  .
\label{x_commutation}%
\end{equation}
If the global vacuum state $\left\vert 0\right\rangle $ is defined by the
condition $\forall\mathbf{k}\widehat{a}\left(  \mathbf{k}\right)  \left\vert
0\right\rangle =0$ then, as we will show below, the field operators create
biorthogonal states.

In the IP the basis vectors are time dependent \cite{CohenQED1}. To
accommodate the possibility of including negative frequency wavefunctions,
$\epsilon=\pm$ states will be defined as%
\begin{align}
\left\vert \phi^{\epsilon}\left(  x\right)  \right\rangle  &  =\sqrt{\hbar
}\int\frac{\mathrm{d}\mathbf{k}}{\left(  2\pi\right)  ^{3}2\omega_{\mathbf{k}%
}}\mathrm{e}^{\mathrm{i}\epsilon\left(  \omega_{\mathbf{k}}t-\mathbf{k}%
\cdot\mathbf{x}\right)  }\left\vert 1_{\mathbf{k}}\right\rangle
,\label{phi_epsilon}\\
\left\vert \pi^{\epsilon}\left(  x\right)  \right\rangle  &  =\sqrt{\hbar}%
\int\frac{\mathrm{d}\mathbf{k}}{\left(  2\pi\right)  ^{3}2}\mathrm{e}%
^{\mathrm{i}\epsilon\left(  \omega_{\mathbf{k}}t-\mathbf{k}\cdot
\mathbf{x}\right)  }\left\vert 1_{\mathbf{k}}\right\rangle ,
\label{pi_epsilon}%
\end{align}
where $\left\vert \pi^{\epsilon}\left(  x\right)  \right\rangle \equiv
c\widehat{D}^{1/2}\left\vert \phi^{\epsilon}\left(  x\right)  \right\rangle $
so that the phase factor $\mathrm{i}$ has been absorbed into the bases and
\begin{equation}
\mathrm{i}\partial_{t}\left\vert \phi^{\epsilon}\left(  x\right)
\right\rangle =-\epsilon c\widehat{D}^{1/2}\left\vert \phi^{\epsilon}\left(
x\right)  \right\rangle . \label{t_dependence}%
\end{equation}
With these definitions annihilation is described by projection onto
$\left\langle \pi^{+}\left(  x\right)  \right\vert $ so that $\left\langle
\pi^{+}\left(  x\right)  |\psi^{+}\right\rangle \ $is positive frequency
while
\begin{equation}
\left\langle \pi^{-}\left(  x\right)  |\psi^{-}\right\rangle =\left\langle
\pi^{+}\left(  x\right)  |\psi^{+}\right\rangle ^{\ast}=\left\langle \psi
^{+}|\pi^{+}\left(  x\right)  \right\rangle \label{negative}%
\end{equation}
is negative frequency where $\epsilon=+$ refers to a particle arriving from
the past and absorbed on $n$, while $\epsilon=-$ refers to a particle emitted
on $n$ and propagating into the future.\textbf{\ }These basis vectors are
biorthogonal in the sense that{\ }%
\begin{equation}
\left\langle \pi^{\epsilon}\left(  x\right)  |\phi^{\epsilon^{\prime}}\left(
y\right)  \right\rangle =\frac{\hbar}{2}{\delta_{n}\left(  x-y\right)  }%
\delta_{\epsilon\epsilon^{\prime}}. \label{x_sp}%
\end{equation}
Based on (\ref{invariant_sp}) there are no $\epsilon=+/\epsilon=-$ cross terms
in the scalar product. The notation ${\delta_{n}\left(  x-y\right)  }$ is
defined to {select }$x$ and $y$ {such} that $x^{\mu}=y^{\mu}$ on the
hyperplane with normal $n_{\mu}.$ Since $\left\vert \phi^{\epsilon}\left(
x\right)  \right\rangle $ and $\left\vert \pi^{\epsilon}\left(  x\right)
\right\rangle $ are biorthogonal, they satisfy the completeness relation%
\begin{equation}
\widehat{1}=\frac{2}{\hbar}\sum_{\epsilon=\pm}\int\mathrm{d}\mathbf{x}%
\left\vert \phi^{\epsilon}\left(  x\right)  \right\rangle \left\langle
\pi^{\epsilon}\left(  x\right)  \right\vert \label{x_complete}%
\end{equation}
where the factor $2/\hbar$ is due to normalization (see (\ref{x_sp})). The
states (\ref{phi_epsilon}) can be interpreted as basis states for solutions to
the first quantized KG wave equation or as one-particle states in QFT.

It can be verified by substitution that the basis states (\ref{phi_epsilon})
and (\ref{pi_epsilon}) are eigenvectors of a position operator of the form
(\ref{F}),%
\begin{equation}
\widehat{\mathbf{x}}=\frac{2}{\hbar}\sum_{\epsilon=\pm}\int\mathrm{d}%
\mathbf{x\ x}\left\vert \phi^{\epsilon}\left(  x\right)  \right\rangle
\left\langle \pi^{\epsilon}\left(  x\right)  \right\vert ,
\label{PositionOperator}%
\end{equation}
and its adjoint, that is%
\begin{align}
\widehat{\mathbf{x}}\left\vert \phi^{\epsilon}\left(  x\right)  \right\rangle
&  =\mathbf{x}\left\vert \phi^{\epsilon}\left(  x\right)  \right\rangle
,\label{eq:XEigen}\\
\widehat{\mathbf{x}}^{\dagger}\left\vert \pi^{\epsilon}\left(  x\right)
\right\rangle  &  =\mathbf{x}\left\vert \pi^{\epsilon}\left(  x\right)
\right\rangle , \label{eq:XDaggarEigen}%
\end{align}
consistent with their biorthogonality. Any classical or one-particle state can
therefore be projected onto the configuration space basis as%
\begin{align}
\left\vert \psi\left(  t\right)  \right\rangle  &  =\widehat{1}\left\vert
\psi\left(  t\right)  \right\rangle \nonumber\\
&  =\frac{2}{\hbar}\sum_{\epsilon=\pm}\int\mathrm{d}\mathbf{x}\left\vert
\phi^{\epsilon}\left(  x\right)  \right\rangle \left\langle \pi^{\epsilon
}\left(  x\right)  |\psi\left(  t\right)  \right\rangle ,\label{x_basis}\\
\left\vert \widetilde{\psi}\left(  t\right)  \right\rangle  &  =\widehat{1}%
\left\vert \widetilde{\psi}\left(  t\right)  \right\rangle \nonumber\\
&  =\frac{2}{\hbar}\sum_{\epsilon=\pm}\int\mathrm{d}\mathbf{x}\left\vert
\pi^{\epsilon}\left(  x\right)  \right\rangle \left\langle \phi^{\epsilon
}\left(  x\right)  |\widetilde{\psi}\left(  t\right)  \right\rangle .
\end{align}
The wave function
\begin{equation}
\psi^{\epsilon}\left(  x\right)  =\left\langle \pi^{\epsilon}\left(  x\right)
|\psi^{\epsilon}\left(  t\right)  \right\rangle \label{wf}%
\end{equation}
completely describes the state $\left\vert \Psi\left(  t\right)  \right\rangle
$ in the $\left\{  \left\vert \phi^{\epsilon}\left(  x\right)  \right\rangle
\right\}  $ basis of position eigenvectors. It may have positive frequency
$\left(  \epsilon=+\right)  $ and negative frequency $\left(  \epsilon
=-\right)  $ components. According to the rules of biorthogonal QM outlined in
Section 1 we have the equality
\begin{equation}
\left\langle \phi^{\epsilon}\left(  x\right)  |\widetilde{\psi}^{\epsilon
}\left(  t\right)  \right\rangle =\left\langle \pi^{\epsilon}\left(  x\right)
|\psi^{\epsilon}\left(  t\right)  \right\rangle . \label{amplitude}%
\end{equation}
Using (\ref{x_sp}), (\ref{x_basis}) and (\ref{amplitude}) the squared norm of
$\left\vert \psi\left(  t\right)  \right\rangle ,$
\begin{equation}
\left\langle \widetilde{\psi}|\psi\right\rangle =\frac{2}{\hbar}\sum
_{\epsilon=\pm}\int\mathrm{d}\mathbf{x}\left\vert \left\langle \pi^{\epsilon
}\left(  x\right)  |\psi^{\epsilon}\left(  t\right)  \right\rangle \right\vert
^{2},
\end{equation}
and the probability density,%
\begin{equation}
p^{\epsilon}\left(  x\right)  =\frac{2}{\hbar}\left\vert \left\langle
\pi^{\epsilon}\left(  x\right)  |\psi^{\epsilon}\left(  t\right)
\right\rangle \right\vert ^{2}, \label{p_KG}%
\end{equation}
is positive definite where $\left\langle \widetilde{\psi}|\psi\right\rangle
=1$ for a one-particle state. The expectation value of the position operator
(\ref{PositionOperator}) is $\left\langle \widetilde{\psi}\left\vert
\widehat{\mathbf{x}}\right\vert \psi\right\rangle =\frac{2}{\hbar}%
\sum_{\epsilon=\pm}\int\mathrm{d}\mathbf{x}\psi^{\epsilon}\left(  x\right)
^{\ast}\mathbf{x}\psi^{\epsilon}\left(  x\right)  .$

Since this application of biorthogonal QM is based on an invariant scalar
product, the QM that it describes is covariant. In particular, the wave
function of a plane wave, $\left\vert 1_{\mathbf{q}}\right\rangle ,$ is
$\left\langle 1_{\mathbf{k}}|1_{\mathbf{q}}\right\rangle =\left(  2\pi\right)
^{3}2\omega_{\mathbf{q}}\,\delta\left(  \mathbf{k}-\mathbf{q}\right)  \ $in
Fourier space and $\left\langle \phi^{\epsilon}\left(  x\right)
|1_{\mathbf{q}}\right\rangle =\sqrt{\hbar}\mathrm{e}^{-\mathrm{i}%
\epsilon\left(  \omega_{\mathbf{k}}t-\mathbf{q}\cdot\mathbf{x}\right)  }$ in
configuration space and a localized state, $\left\vert \phi^{\epsilon}\left(
y\right)  \right\rangle ,$ is $\left\langle 1_{\mathbf{k}}|^{\epsilon}%
\phi\left(  y\right)  \right\rangle =\sqrt{\hbar}\mathrm{e}^{\mathrm{i}%
\epsilon\left(  \omega_{\mathbf{k}}t-\mathbf{k}\cdot\mathbf{y}\right)  }$ in
Fourier space and $\left\langle \pi^{\epsilon}\left(  x\right)  |\phi
^{\epsilon}\left(  y\right)  \right\rangle =\frac{\hbar}{2}{\delta_{n}\left(
x-y\right)  }$ in configuration space. Eqs. (\ref{x_complete}) and
(\ref{PositionOperator}) can be generalized to%
\begin{align}
\widehat{1}  &  =\frac{2}{\hbar}\int\mathrm{d}\sigma\left\vert \phi\left(
x\right)  \right\rangle \left\langle -\mathrm{i}\epsilon n_{\mu}\partial^{\mu
}\phi\left(  x\right)  \right\vert ,\label{invariant_1}\\
\widehat{x}_{i}  &  =\frac{2}{\hbar}\int\mathrm{d}\sigma x_{i}\left\vert
\phi\left(  x\right)  \right\rangle \left\langle -\mathrm{i}\epsilon n_{\mu
}\partial^{\mu}\phi\left(  x\right)  \right\vert \label{invariant_x}%
\end{align}
respectively where $x_{\mu}n^{\mu}=ct_{0}$ on the hyperplane with normal
$n_{\mu}$ and $t_{0}$ is the hyperplane of simultaneity
\cite{Fleming,HawtonPhysicaScripta}. The matrix representing the position
observable in configuration space is
\begin{equation}
x_{i}\left(  x,y\right)  =\sum_{\epsilon=\pm}\left\langle \pi^{\epsilon
}\left(  x\right)  \left\vert \widehat{x}_{i}\right\vert \phi^{\epsilon
}\left(  y\right)  \right\rangle _{n}=x_{i}{\delta_{n}\left(  x-y\right)  }%
\end{equation}
where $x_{i}$ is on the $n$ hyperplane.

The relationship of the relativistic position operator to the nonrelativistic
position operator $\mathrm{i}\mathbf{\nabla}_{\mathbf{k}}$ and the NW position
operator can be seen by transforming to Fourier space. The position operator
(\ref{PositionOperator}) is in the IP while the conventional position operator
is time independent so it is in the Schr\"{o}dinger picture (SP). For positive
frequency fields the SP position operator (\ref{PositionOperator}) is
\begin{equation}
\widehat{\mathbf{x}}^{{\mathrm{SP}}}=\int\frac{\mathrm{d}\mathbf{k}}{2\left(
2\pi\right)  ^{3}}\int\frac{\mathrm{d}\mathbf{q}}{2\left(  2\pi\right)  ^{3}%
}\int{\mathrm{d}\mathbf{x\ x}\,\mathrm{e}}^{-\mathrm{i}\mathbf{k}%
\cdot\mathbf{x}}\left\vert \frac{1_{\mathbf{k}}}{\omega_{\mathbf{k}}%
}\right\rangle \left\langle 1_{\mathbf{q}}\right\vert {\mathrm{e}}%
^{\mathrm{i}\mathbf{q}\cdot\mathbf{x}}.
\end{equation}
Since $\int{\mathrm{d}\mathbf{x}}\,{\mathrm{e}}^{\mathrm{i}\mathbf{q}%
\cdot\mathbf{x}}\mathrm{i}\mathbf{\nabla}_{\mathbf{k}}{\mathrm{e}%
}^{-\mathrm{i}\mathbf{k}\cdot\mathbf{x}}=2\left(  2\pi\right)  ^{3}%
\mathrm{i}\mathbf{\nabla}_{\mathbf{k}}\delta\left(  \mathbf{q}-\mathbf{k}%
\right)  $ \cite{Sakurai},%
\begin{align}
\widehat{\mathbf{x}}^{{\mathrm{SP}}}  &  =\int\frac{\mathrm{d}\mathbf{k}%
}{2\left(  2\pi\right)  ^{3}}\left\vert \frac{1_{\mathbf{k}}}{\omega
_{\mathbf{k}}}\right\rangle \mathrm{i}\mathbf{\nabla}_{\mathbf{k}}\left\langle
1_{\mathbf{k}}\right\vert ,\\
\widehat{\mathbf{x}}^{{\mathrm{SP}}}\left(  \mathbf{k}\right)   &
=\mathrm{i}\mathbf{\nabla}_{\mathbf{k}}, \label{pos_op}%
\end{align}
so that $\mathrm{i}\mathbf{\nabla}_{\mathbf{k}}$ is the position operator in
the $\left\vert 1_{\mathbf{k}}/\omega_{\mathbf{k}}\right\rangle \left\langle
1_{\mathbf{k}}\right\vert $ basis. When operating on $\mathrm{e}%
^{-\mathrm{i}\mathbf{k}\cdot\mathbf{x}}$ it extracts the position $\mathbf{x}$
where the particle was created. With positive definite Hermitian metric
$\Theta,$ the scalar product is $\left\langle .|\Theta.\right\rangle $ and an
operator $\widehat{O}$ is quasi-Hermitian if $\Theta\widehat{O}^{\dagger
}=\widehat{O}\Theta$ \cite{GeyerScholtz}. Since according to
(\ref{PositionOperator}) $\widehat{\mathbf{x}}^{\dagger}=\widehat{D}%
^{1/2}\widehat{\mathbf{x}}\widehat{D}^{-1/2}$, the position operator is
quasi-Hermitian. Indeed defining $S\equiv\Theta^{1/2},$ the operator
$\widehat{o}=\widehat{o}^{\dagger}=S^{-1}\widehat{O}^{\dagger}S$ is Hermitian.
In (\ref{k_sp}) the metric is $\Theta=\omega_{\mathbf{k}}^{-1}$ so
$S=\omega_{\mathbf{k}}^{-1/2},$ the basis is $\left\{  \omega_{\mathbf{k}%
}^{-1/2}\left\vert 1_{\mathbf{k}}\right\rangle \right\}  $ and the matrix
representing the NW position operator
\cite{Mostafazadeh1,NewtonWigner,Mostafazadeh2},%
\begin{equation}
\widehat{\mathbf{x}}_{{\mathrm{NW}}}^{{\mathrm{SP}}}\left(  \mathbf{k}\right)
=\omega_{\mathbf{k}}^{1/2}\mathrm{i}\mathbf{\nabla}_{\mathbf{k}}%
\omega_{\mathbf{k}}^{-1/2},
\end{equation}
is of the form $\widehat{o}^{\dagger}=S^{-1}\widehat{O}^{\dagger}S$ for
$\widehat{O}^{\dagger}=\widehat{\mathbf{x}}^{{\mathrm{SP}}}\left(
\mathbf{k}\right)  $ where $\omega_{\mathbf{k}}^{1/2}\mathrm{i}\mathbf{\nabla
}_{\mathbf{k}}\omega_{\mathbf{k}}^{-1/2}=\mathrm{i}\mathbf{\nabla}%
_{\mathbf{k}}-\frac{\mathrm{i}c^{2}}{\omega_{\mathbf{k}}^{2}}\mathbf{k}$. The
factors $\omega_{\mathbf{k}}^{\pm1/2}$ introduce nonlocality into the
configuration space description of the position eigenvectors. This nonlocality
is not physical since it does not appear in the manifestly covariant
description of the position observable. This is consistent with the metric
operator being not physically observable as discussed in \cite{Brody2}.

\section{Photons}

For photons the scalar field $\phi$ should be replaced with the four-vector
potential $A^{\mu}$. Both $A_{\mu}\partial^{\nu}A^{\mu}=\left(  A_{\mu
}\partial_{ct}A^{\mu},A_{\mu}\nabla A^{\mu}\right)  $ and $A_{\mu}F^{\mu\nu
}=\left(  \mathbf{A}\cdot\mathbf{E}/c,\mathbf{A}\times\mathbf{B}\right)  $
multiplied by $\mathrm{i}\epsilon_{0}c/\hbar$ are candidates for the
four-current density, $F^{\mu\nu}=\partial^{\mu}A^{\nu}-\partial^{\nu}%
A^{\mu\nu}$ being the Faraday tensor. The properties of an operator of the
form $\mathrm{i}A_{\nu}F^{\nu\mu}$ were investigated in \cite{NumDens95}. The
Coulomb gauge condition $\nabla\cdot\mathbf{A}=0$ is not Lorentz invariant,
but $A_{\mu}$ can be chosen to transform as a Lorentz four-vector up to an
extra term that maintains the Coulomb gauge in all frames of reference
\cite{Debierre}. To avoid the complications associated with nonphysical
longitudinal and scalar photons the Coulomb gauge will be used in this
Section. In a source-free region in the Coulomb gauge both $A_{\mu}%
\partial^{\nu}A^{\mu}$ and $A_{\nu}F^{\nu\mu}$ reduce to $\left(
\mathbf{A}_{\perp}\cdot\mathbf{E}_{\perp}/c,\mathbf{A}_{\perp}\times
\mathbf{B}\right)  $.

Following (\ref{M}) we can define%
\begin{equation}
J^{\mu}\left(  x\right)  =\frac{\mathrm{i}\epsilon_{0}c}{\hbar}\sum
_{\lambda,i}A_{\lambda i}\left(  x\right)  ^{\ast}\overleftrightarrow{\partial
}^{\mu}A_{\lambda ci}\left(  x\right)  \label{Jph}%
\end{equation}
where\textbf{\ }$\mathbf{A}_{\lambda}$\textbf{\ }is a transverse vector
potential of helicity\textbf{\ }$\lambda$ that satisfies the classical Maxwell
wave equation {$\left(  \widehat{D}+\partial_{ct}^{2}\right)  \mathbf{A}%
_{\lambda}=0$ }where\textbf{\ }$m=0$\textbf{\ }so{\ $\widehat{D}=-\nabla^{2}.$
The} conjugate field is\textbf{\ }$\mathbf{A}_{\lambda c}\equiv\mathrm{i}%
\widehat{D}^{-1/2}\partial_{ct}\mathbf{A}_{\lambda}=\mathbf{A}_{\lambda}%
^{+}-\mathbf{A}_{\lambda}^{-}$ where\textbf{\ }$\mathbf{A}_{\lambda
}=\mathbf{A}_{\lambda}^{+}+\mathbf{A}_{\lambda}^{-}.$\textbf{\ }{It can then
be verified by substitution that (\ref{Jph}) satisfies the continuity equation
$\partial_{\mu}J^{\mu}\left(  x\right)  =0$. As a consequence the scalar
product $\int_{\sigma}\mathrm{d}\sigma n_{\mu}J^{\mu}\left(  x\right)  $ is
Lorentz invariant up to a term that maintains the Coulomb gauge.
}The{{\ photon scalar product that replaces (\ref{sp}) is%
\begin{equation}
\left(  A_{1},A_{2}\right)  =\frac{2\epsilon_{0}c}{\hbar}\sum_{\lambda
,\epsilon,i}\left\langle A_{1\lambda i}^{\epsilon}|\widehat{D}^{1/2}%
A_{2\lambda i}^{\epsilon}\right\rangle \label{photon_sp}%
\end{equation}
where }$A=A^{\mu}=\left(  0,\mathbf{A}_{\perp}\right)  $. }

For photons described in the Coulomb gauge, the field operator is
$\widehat{\mathbf{A}}_{\perp}\left(  \mathbf{x},t\right)  $ and its canonical
conjugate is $-\epsilon_{0}\widehat{\mathbf{E}}_{\perp}\left(  \mathbf{x}%
,t\right)  $ where $\widehat{\mathbf{E}}_{\perp}=-\partial_{t}%
\widehat{\mathbf{A}}_{\perp} $ is the electric field operator. The Fourier
space spherical polar coordinates will be called $k,$ $\theta_{\mathbf{k}}$
and $\phi_{\mathbf{k}}$ and their corresponding unit vectors $\mathbf{e}%
_{\mathbf{k}},$ $\mathbf{e}_{\theta}$ and $\mathbf{e}_{\phi}.$ The definite
helicity transverse unit vectors are $\mathbf{e}_{\lambda}\left(
\mathbf{k}\right)  =\left(  \mathbf{e}_{\theta}+\mathrm{i}\lambda
\mathbf{e}_{\phi}\right)  /\sqrt{2}$ where $\lambda$ is helicity.{\ }The NW
photon position operator with commuting components can be written in Fourier
space as \cite{HawtonBaylis} $\widehat{\mathbf{x}}=\widehat{E}\left(
\mathrm{i}\omega_{\mathbf{k}}^{1/2}\nabla_{\mathbf{k}}\omega_{\mathbf{k}%
}^{-1/2}\right)  \widehat{E}^{-1}$ where $\widehat{E}$ is a rotation through
Euler angles to fixed reference axes. The basic idea is the same as was used
in the derivation of the NW position operator in \cite{Mostafazadeh1}; the
position information is contained in the factor $\exp\left(  \mathrm{i}%
\mathbf{k\cdot x}\right)  $ in the wave function but the factor $\omega
_{\mathbf{k}}^{1/2}$ must be eliminated before the nonrelativistic position
operator $\mathrm{i}\nabla_{\mathbf{k}}$ can be used to extract this
information. For the transverse fields that describe photons an additional
unitary transformation $\widehat{E}$ that rotates the field vectors to axes
fixed in space is needed. A position eigenvector has a vortex structure like
twisted light \cite{Twisted,HawtonBaylis} in which the photon position
eigenvalue $\mathbf{x}$ is the center of internal angular momentum
\cite{Fleming}. We have seen here that it is also the parameter $\mathbf{x}$
in the field operators. In the position operator derived in \cite{HawtonPosOp}
the relevant case is $\alpha=0.$

Following the derivation in Section 3, the IP photon position operator is%
\begin{equation}
\widehat{\mathbf{x}}=\frac{2}{\hbar}\sum_{\epsilon,\lambda=\pm}\int%
\mathrm{d}\mathbf{x\ x}\left\vert \mathbf{A}_{\lambda}^{\epsilon}\left(
x\right)  \right\rangle \cdot\left\langle \mathbf{E}_{\lambda}^{\epsilon
}\left(  x\right)  \right\vert ,\label{x_photon}%
\end{equation}
the position eigenvector equations are%
\begin{align}
\widehat{\mathbf{x}}\left\vert \mathbf{A}_{\lambda}^{\epsilon}\left(
x\right)  \right\rangle  &  =\mathbf{x}\left\vert \mathbf{A}_{\lambda
}^{\epsilon}\left(  x\right)  \right\rangle ,\label{photon_evecs}\\
\widehat{\mathbf{x}}^{\dagger}\left\vert \mathbf{E}_{\lambda}^{\epsilon
}\left(  x\right)  \right\rangle  &  =\mathbf{x}\left\vert \mathbf{E}%
_{\lambda}^{\epsilon}\left(  x\right)  \right\rangle
\end{align}
and position basis states are%
\begin{align}
\left\vert \mathbf{A}_{\lambda}^{+}\left(  x\right)  \right\rangle  &
\equiv\widehat{\mathbf{A}}_{\lambda}^{-}\left(  x\right)  \left\vert
0\right\rangle ,\label{A_loc}\\
\left\vert \mathbf{A}_{\lambda}^{-}\left(  x\right)  \right\rangle  &
=\left\vert \mathbf{A}_{\lambda}^{+}\left(  x\right)  \right\rangle ^{\ast},\\
\left\vert \mathbf{E}_{\lambda}^{\epsilon}\left(  x\right)  \right\rangle  &
\equiv c\widehat{D}^{1/2}\left\vert \mathbf{A}_{\lambda}^{\epsilon}\left(
x\right)  \right\rangle .\label{E_loc}%
\end{align}
Here the potential operator reads
\begin{equation}
\widehat{\mathbf{A}}_{\lambda}^{-}\left(  x\right)  =\sqrt{\frac{\hbar
}{\epsilon_{0}}}\int\frac{\mathrm{d}\mathbf{k}}{\left(  2\pi\right)
^{3}2\omega_{\mathbf{k}}}\mathbf{e}_{\lambda}\left(  \mathbf{k}\right)
\mathrm{e}^{-\mathrm{i}\left(  \mathbf{k}\cdot\mathbf{x}-\omega_{\mathbf{k}%
}t\right)  }\hat{a}_{\lambda}^{\dagger}\left(  \mathbf{k}\right)  .\label{A}%
\end{equation}
The Fourier space canonical commutation relations and orthogonality relations
are%
\begin{align}
\left[  \widehat{a}_{\lambda}\left(  \mathbf{k}\right)  ,\widehat{a}_{\sigma
}^{\dagger}\left(  \mathbf{q}\right)  \right]    & =\left(  2\pi\right)
^{3}2\omega_{\mathbf{k}}\,\delta\left(  \mathbf{k}-\mathbf{q}\right)
\delta_{\lambda\sigma},\label{photon_k_commutation}\\
\left\langle 1_{\lambda,\mathbf{k}}|1_{\sigma,\mathbf{q}}\right\rangle  &
=\left(  2\pi\right)  ^{3}2\omega_{\mathbf{k}}\,\delta\left(  \mathbf{k}%
-\mathbf{q}\right)  \delta_{\lambda\sigma},\label{photon_k_normalization}%
\end{align}
where $\left\vert 1_{\lambda,\mathbf{k}}\right\rangle \equiv a_{\lambda
}^{\dagger}\left(  \mathbf{k}\right)  \left\vert 0\right\rangle .$ These
photon position eigenvectors are biorthogonal that is
\begin{equation}
\sum_{i=1}^{3}\left\langle E_{\lambda i}^{\epsilon}\left(  x\right)
|A_{\sigma i}^{\epsilon^{\prime}}\left(  y\right)  \right\rangle =\frac{\hbar
}{2\epsilon_{0}}{\delta_{n}\left(  x-y\right)  }\delta_{\lambda\sigma}%
\delta_{\epsilon\epsilon^{\prime}}\label{orthogonal}%
\end{equation}
where the subscripts $i$ denote Cartesian components of the
three-vectors,\textbf{\ }$\epsilon=+$ for absorption at $x$ while $\epsilon=-$
for emission at $x,$ and the scalar product (\ref{photon_sp}) is\textbf{\ }%
$\left(  A_{\lambda}^{\epsilon}\left(  x\right)  ,A_{\sigma}^{\epsilon}\left(
y\right)  \right)  ={\delta_{n}\left(  x-y\right)  }\delta_{\lambda\sigma}.$
For free photons described by transverse fields the completeness relation is%
\begin{equation}
\widehat{1}_{\perp}=\frac{2\epsilon_{0}}{\hbar}\sum_{\epsilon,\lambda=\pm}%
\int\mathrm{d}\mathbf{x}\left\vert \mathbf{A}_{\lambda}^{\epsilon}\left(
x\right)  \right\rangle \cdot\left\langle \mathbf{E}_{\lambda}^{\epsilon
}\left(  x\right)  \right\vert \label{complete}%
\end{equation}
where we have defined%
\begin{equation}
\left\vert \mathbf{A}_{\lambda}^{\epsilon}\left(  x\right)  \right\rangle
\cdot\left\langle \mathbf{E}_{\lambda}^{\epsilon}\left(  x\right)  \right\vert
\equiv\sum_{i=1}^{3}\left\vert A_{\lambda i}^{\epsilon}\left(  x\right)
\right\rangle \left\langle E_{\lambda i}^{\epsilon}\left(  x\right)
\right\vert .\label{dot}%
\end{equation}
The identity operator $\widehat{1}_{\perp}$ on the space of transverse photons
is closely connected to the so-called `transverse Dirac delta' of QED
\cite{CohenQED1}. For any transverse state we can thence write
\begin{equation}
\left\vert \psi_{\perp}\left(  t\right)  \right\rangle =\frac{2\epsilon_{0}%
}{\hbar}\sum_{\epsilon,\lambda=\pm}\int\mathrm{d}\mathbf{x}\left\vert
\mathbf{A}_{\lambda}^{\epsilon}\left(  x\right)  \right\rangle \cdot
\left\langle \mathbf{E}_{\lambda}^{\epsilon}\left(  x\right)  |\psi\left(
t\right)  \right\rangle \label{one_photon}%
\end{equation}
and the wave function
\begin{equation}
\mathbf{\psi}_{\lambda}^{\epsilon}\left(  x\right)  =\left\langle
\mathbf{E}_{\lambda}^{\epsilon}\left(  x\right)  |\psi\left(  t\right)
\right\rangle =\left\langle \mathbf{A}_{\lambda}^{\epsilon}\left(  x\right)
|\widetilde{\psi}\left(  t\right)  \right\rangle \label{wave_function}%
\end{equation}
completely describes the state $\left\vert \psi_{\perp}\left(  t\right)
\right\rangle $ in either basis of position eigenvectors. The dual state
vector is
\begin{equation}
\left\vert \widetilde{\psi}_{\perp}\left(  t\right)  \right\rangle
=\frac{2\epsilon_{0}}{\hbar}\sum_{\epsilon,\lambda=\pm}\int\mathrm{d}%
\mathbf{x}\left\vert \mathbf{E}_{\lambda}^{\epsilon}\left(  x\right)
\right\rangle \cdot\left\langle \mathbf{A}_{\lambda}^{\epsilon}\left(
x\right)  |\widetilde{\psi}\left(  t\right)  \right\rangle ,
\end{equation}
the squared norm is
\begin{equation}
\left\langle \psi_{\perp}|\widetilde{\psi}_{\perp}\right\rangle =\frac
{2\epsilon_{0}}{\hbar}\sum_{\epsilon,\lambda=\pm}\int\mathrm{d}\mathbf{x}%
\left\vert \mathbf{\psi}_{\lambda}^{\epsilon}\left(  x\right)  \right\vert
^{2}%
\end{equation}
and the \emph{probability density} for a transition from $\left\vert
\psi_{\perp}\left(  t\right)  \right\rangle $ to the $\epsilon$-frequency
position eigenvector with helicity $\lambda$%
\begin{align}
p_{\lambda}^{\epsilon}\left(  x\right)   &  =\frac{2\epsilon_{0}}{\hbar
}\left\vert \mathbf{\psi}_{\lambda}^{\epsilon}\left(  x\right)  \right\vert
^{2}\label{probability}\\
&  =\frac{\left\langle \psi_{\perp}|\widetilde{\psi}_{\perp}\right\rangle
\left\vert \mathbf{\psi}_{\lambda}^{\epsilon}\left(  x\right)  \right\vert
^{2}}{\sum_{\epsilon,\lambda=\pm}\int\mathrm{d}\mathbf{x}\left\vert
\mathbf{\psi}_{\lambda}^{\epsilon}\left(  x\right)  \right\vert ^{2}}\nonumber
\end{align}
is positive definite with $\left\langle \psi|\widetilde{\psi}\right\rangle =1$
for a one-photon state.

A state can be Fourier expanded as%
\begin{equation}
\left\vert \psi_{\perp}\left(  t\right)  \right\rangle =\sum_{\lambda
,\epsilon=\pm}\int\frac{\mathrm{d}\mathbf{k}}{\left(  2\pi\right)  ^{3}%
2\omega_{\mathbf{k}}}c_{\lambda}^{\epsilon}\left(  \mathbf{k},t\right)
\left\vert 1_{\lambda,\mathbf{k}}\right\rangle . \label{k_state}%
\end{equation}
Eqs. (\ref{A_loc}) to (\ref{photon_k_normalization}) give\textbf{\ }%
$\left\langle \mathbf{E}_{\lambda}\left(  \mathbf{x}\right)  |1_{\lambda
,\mathbf{k}}\right\rangle /\omega_{\mathbf{k}}=\left\langle \mathbf{A}%
_{\lambda}\left(  \mathbf{x}\right)  |1_{\lambda,\mathbf{k}}\right\rangle $ so
that $\left\vert 1_{\lambda,\mathbf{k}}/\omega_{\mathbf{k}}\right\rangle
\in\mathcal{H}$\ and\textbf{\ }$\left\vert 1_{\lambda,\mathbf{k}}\right\rangle
\in\mathcal{H}^{\ast}.$\textbf{\ }Substitution in (\ref{wave_function}) then
gives the dual state vector%
\begin{equation}
\left\vert \widetilde{\psi}_{\perp}\left(  t\right)  \right\rangle
=\sum_{\lambda,\epsilon=\pm}\int\frac{\mathrm{d}\mathbf{k}}{\left(
2\pi\right)  ^{3}2}c_{\lambda}^{\epsilon}\left(  \mathbf{k},t\right)
\left\vert 1_{\lambda,\mathbf{k}}\right\rangle . \label{k_dual}%
\end{equation}
The probability amplitude for a transition to a $\epsilon$-frequency plane
wave state with wave vector\textbf{\ }$\mathbf{k}$ and helicity\textbf{\ }%
$\lambda$\textbf{\ }is proportional to\textbf{\ }$\left\langle 1_{\lambda
,\mathbf{k}}|\psi_{\perp}^{\epsilon}\left(  t\right)  \right\rangle
=\left\langle 1_{\lambda,\mathbf{k}}/\omega_{\mathbf{k}}|\widetilde{\psi
}_{\perp}^{\epsilon}\left(  t\right)  \right\rangle =c_{\lambda}^{\epsilon
}\left(  \mathbf{k},t\right)  .$ According to the rules of biorthogonal QM
outlined in the Introduction the probability density for this transition is%
\begin{align}
p_{\lambda}^{\epsilon}\left(  \mathbf{k}\right)   &  =\frac{\left\vert
\left\langle 1_{\lambda,\mathbf{k}}|\psi^{\epsilon}\left(  t\right)
\right\rangle \right\vert ^{2}}{\left(  2\pi\right)  ^{3}2}\nonumber\\
&  =\frac{\left\langle \widetilde{\psi}\left(  t\right)  |\psi\left(
t\right)  \right\rangle \left\vert c_{\lambda}^{\epsilon}\left(
\mathbf{k},t\right)  \right\vert ^{2}}{\sum_{\lambda,\epsilon=\pm}%
\int\mathrm{d}\mathbf{k}\,\left\vert c_{\lambda}^{\epsilon}\left(
\mathbf{k},t\right)  \right\vert ^{2}}. \label{p_k}%
\end{align}
Time dependence of\textbf{\ }$c_{\lambda}^{\epsilon}\left(  \mathbf{k}%
,t\right)  $ indicates the presence of a source. When a photon is emitted by
an atom, the expectation value of the photon number is smaller than one and
approaches unity as $t\rightarrow\infty.$\textbf{\ }If\textbf{\ }$\left\vert
\psi_{\perp}\left(  t\right)  \right\rangle $ is normalized so that\textbf{\ }%
$n\left(  t\right)  =\left\langle \widetilde{\psi}_{\perp}\left(  t\right)
|\psi_{\perp}\left(  t\right)  \right\rangle $ is the number of photons, the
probability density for\textbf{\ }$k^{\mu}=\left(  \epsilon\omega_{\mathbf{k}%
},\mathbf{k}\right)  $ is\textbf{\ }$\frac{1}{\left(  2\pi\right)  ^{3}%
2}\left\vert c_{\lambda}^{\epsilon}\left(  \mathbf{k},t\right)  \right\vert
^{2}$\textbf{\ } while the probability density to find a\textbf{\ }%
photon\textbf{\ }at \textbf{$x$ }on the hyperplane $\sigma$ is $\frac
{2\epsilon_{0}}{\hbar}\left\vert \mathbf{\psi}_{\lambda}^{\epsilon}\left(
x\right)  \right\vert ^{2}$. In the SP the photon position operator
(\ref{x_photon}) can be written as%
\begin{equation}
\widehat{\mathbf{x}}^{{\mathrm{SP}}}\left(  \mathbf{k}\right)  =\widehat{E}%
\mathrm{i}\mathbf{\nabla}_{\mathbf{k}}\widehat{E}^{-1}.
\end{equation}

The scalar product\textbf{\ }$\left\langle \mathbf{E}_{\lambda}^{+}\left(
x\right)  |\psi\left(  t\right)  \right\rangle =\left\langle \mathbf{E}%
_{\lambda}\left(  x\right)  |\psi\left(  t\right)  \right\rangle $ that leads
to an invariant probability to annihilate a photon is proportional to
probability amplitude, not the electric field. {Glauber defined an ideal
photon detector as a system of negligible size with a frequency-independent
photon absorption probability \cite{Glauber}. For the positive frequency
one-photon state $\left\vert \psi\left(  t\right)  \right\rangle $ he found
that the probability to count a photon is proportional to }$\left\vert
\left\langle \mathbf{E}_{\lambda}\left(  x\right)  |\psi\left(  t\right)
\right\rangle \right\vert ^{2}${$.{\ }$Glauber considered photodetection to be
a square law process and interpreted it to be responsive to the density of
electromagnetic energy, but }number density gives an invariant probability to
count a photon while energy density does not.{\ Indeed, the biorthogonal
completeness relation (\ref{complete}) implies that a basis of ideal Glauber
detectors can be defined provided the state vector $\left\vert \psi
\right\rangle \in\mathcal{H}$ of the photon at hand has been created by the
$\mathbf{A}\cdot\mathbf{p}$ minimal coupling Hamiltonian. In that case,
$\left\vert \left\langle \mathbf{E}_{\lambda}\left(  x\right)  |\psi\left(
t\right)  \right\rangle \right\vert ^{2}$ is proportional to photon
probability density.}

{\ } Here a positive definite particle density is obtained in the physical
Hilbert space according to the rules of biorthogonal QM summarized in the
Introduction. An alternative approach is to transform the physical fields
to{\ the NW representation using the nonlocal operator }$\widehat{D}^{-1/4}$
and its inverse{\ {\cite{Mostafazadeh1,BabaeiMostafazadeh}}. The disadvantage
to this approach is that the relationship between the NW wave function and a
current source is nonlocal. Here the field due to the local interaction
Hamilton }$\widehat{J}_{\mu}\left(  x\right)  \widehat{A}^{\mu}\left(
x\right)  ${\ is calculated first and the probability amplitude for a
transition to a position eigenvector is then obtained using the invariant
scalar product (\ref{photon_sp}). These fields have well defined Lorentz and
gauge transformation properties. The positive definiteness of the probability
follows then directly from the mathematical rules of biorthogonal QM.}

An advantage of the second quantized formalism included here is that
multiphoton wave functions can be introduced as in
\cite{Glauber,SmithRaymer,Hawton07}. For example, to the two photon state
$\left\vert \psi_{2}\right\rangle $ can be associated the wave function%
\begin{equation}
\mathbf{\psi}_{\lambda_{1},\lambda_{2}}\left(  \mathbf{x}_{1},\mathbf{x}%
_{2},t\right)  =\left\langle \mathbf{E}_{\lambda_{1}}\left(  \mathbf{x}%
_{1}\right)  \mathbf{E}_{\lambda_{2}}\left(  \mathbf{x}_{2}\right)  |\psi
_{2}\left(  t\right)  \right\rangle
\end{equation}
with $\left\vert \mathbf{E}_{\lambda_{1}}\left(  \mathbf{x}_{1}\right)
\mathbf{E}_{\lambda_{2}}\left(  \mathbf{x}_{2}\right)  \right\rangle
\equiv\widehat{\mathbf{E}}_{\lambda_{1}}\left(  \mathbf{x}_{1}\right)
\widehat{\mathbf{E}}_{\lambda_{2}}\left(  \mathbf{x}_{2}\right)  \left\vert
0\right\rangle .$ This wave function localizes the photons at spatial points
$\mathbf{x}_{1}$ and $\mathbf{x}_{2}$ at time $t$ and can describe entangled
two-photon states.

\section{Wave function of a photon emitted by an atom}

The wave function (\ref{wave_function}) for a photon emitted by a two-level
atom initially in its excited state was derived in \cite{Sipe} and
\cite{Debierre,DebierreDurt}, to{\ first order in the IP minimal coupling
interaction Hamiltonian $\widehat{H}_{I}=\left(  e/m_{e}\right)
\widehat{\mathbf{A}}\left(  \hat{\mathbf{x}}_{e},t\right)  \cdot
\hat{\mathbf{p}}_{e}$}$\left(  t\right)  .$ For a two-level atom initially in
its excited state $\left\vert \mathrm{e}\right\rangle $ with no photons
present, the positive frequency IP wave function describing decay to its
ground state $\left\vert \mathrm{g}\right\rangle $ while emitting one photon
is%
\begin{align}
\left\vert \psi\left(  t\right)  \right\rangle  &  =c_{\mathrm{e}}\left(
t\right)  \left\vert \mathrm{e},0\right\rangle \label{eq:IPState}\\
&  +\sum_{\lambda=\pm}\int\frac{\mathrm{d}\mathbf{k}}{\left(  2\pi\right)
^{3}2\omega_{\mathbf{k}}}c_{\mathrm{g},\lambda}\left(  \mathbf{k},t\right)
\left\vert \mathrm{g},1_{\lambda,\mathbf{k}}\right\rangle \nonumber
\end{align}
where
\begin{align}
c_{\mathrm{g},\lambda}\left(  \mathbf{k},t\right)   &  =\frac{e}{m_{e}%
}\left\langle \mathrm{g},1_{\lambda,\mathbf{k}}\left\vert \widehat{\mathbf{A}%
}_{\lambda}^{-}\left(  \hat{\mathbf{x}}_{e},t\right)  \cdot\hat{\mathbf{p}%
}_{e}\left(  t\right)  \right\vert \mathrm{e},0\right\rangle
\label{eq:GroundPhoton}\\
&  \times\frac{1-\mathrm{e}^{\mathrm{i}\left(  \omega_{\mathbf{k}}-\omega
_{0}\right)  t}}{\hbar\left(  \omega_{\mathbf{k}}-\omega_{0}\right)
}.\nonumber
\end{align}
Here $\hbar\omega_{0}$ is the level separation between the ground and excited
states and $\hat{\mathbf{x}}_{e}$ and $\hat{\mathbf{p}}_{e}$ are the electron
position and momentum operators. For $\left\vert \widetilde{\psi}_{\perp
}\right\rangle $ given by (\ref{k_dual}) the transverse single-photon state
and its dual are thus given by \label{eq:PsiPsiTilde}
\begin{align}
\left\vert \psi_{\perp}\left(  t\right)  \right\rangle  &  =\sum_{\lambda=\pm
}\int\frac{\mathrm{d}\mathbf{k}}{\left(  2\pi\right)  ^{3}2\omega_{\mathbf{k}%
}}c_{\mathrm{g},\lambda}\left(  \mathbf{k},t\right)  \left\vert 1_{\lambda
,\mathbf{k}}\right\rangle ,\label{phi&psitilde}\\
\left\vert \widetilde{\psi}_{\perp}\left(  t\right)  \right\rangle  &
=\sum_{\lambda=\pm}\int\frac{\mathrm{d}\mathbf{k}}{\left(  2\pi\right)  ^{3}%
2}c_{\mathrm{g},\lambda}\left(  \mathbf{k},t\right)  \left\vert 1_{\lambda
,\mathbf{k}}\right\rangle .
\end{align}

In \cite{Debierre,DebierreDurt} the minimal coupling Hamiltonian created the
photon state vector $\left\vert \psi_{\perp}\right\rangle $ in the $\left\vert
\mathbf{A}_{\lambda}\left(  \mathbf{x}\right)  \right\rangle $ basis so the
appropriate wave function is $\left\langle \mathbf{E}_{\lambda}^{\epsilon
}\left(  \mathbf{x}\right)  |\psi_{\perp}\right\rangle $. Indeed we have from
(\ref{wave_function}) and {(\ref{eq:IPState})}
\begin{equation}
\mathbf{\psi}_{\lambda}^{+}\left(  x\right)  =\mathrm{i}\sqrt{\frac{\hbar
}{\epsilon_{0}}}\int\frac{\mathrm{d}\mathbf{k}}{\left(  2\pi\right)  ^{3}%
2}\mathbf{e}_{\lambda}\left(  \mathbf{k}\right)  \mathrm{e}^{-\mathrm{i}%
k_{\mu}x^{\mu}}c_{\mathrm{g},\lambda}\left(  \mathbf{k},t\right)  .
\label{atom_wf}%
\end{equation}
The positive frequency wave function of the emitted photon is calculated, but
causal solutions that include negative frequencies are also considered. Taking
into account a factor $-\mathrm{i}$ in the electric field operator used in
\cite{Debierre,DebierreDurt}, $c_{\mathrm{g},\lambda}=-\mathrm{i}c_{\lambda
}^{+}$ in (\ref{k_state})\textbf{.} Substitution of the wave function
$\mathbf{\psi}_{\lambda}\left(  x\right)  =\left\langle \mathbf{E}_{\lambda
}\left(  \mathbf{x}\right)  |\psi_{\perp}\left(  t\right)  \right\rangle $ in
(\ref{probability}) gives the probability density in space to count a photon
at time $t.$\textbf{\ }Since in \cite{Debierre,DebierreDurt} the wave function
is normalized as $\left\langle \psi_{\perp}|\psi_{\perp}\right\rangle =1,$ the
factor $\left\langle \psi_{\perp}|\widetilde{\psi}_{\perp}\right\rangle $ in
(\ref{probability}) and (\ref{p_k}) approaches\textbf{\ }$1/\omega_{0}$ as $t$
$\rightarrow\infty.$

If the standard (dipolar) $\mathbf{E}\cdot\mathbf{x}$ Hamiltonian were to be
used instead, the photon would be created in the $\left\vert \mathbf{E}%
_{\lambda}\left(  x\right)  \right\rangle $\ basis and the appropriate wave
function would be $\left\langle \mathbf{A}_{\lambda}\left(  x\right)
|\widetilde{\psi}_{\perp}\left(  t\right)  \right\rangle .$

\section{Localized states}

In Sections 3 and 4, starting with biorthogonal bases motivated by the
canonical commutation relations, a bottom up version of the formalism of
biorthogonal QM \cite{Brody,Brody2} was used to derive relativistic QM for KG
particles and photons. For generality, a scalar product that is positive
definite for both positive and negative frequency fields \cite{Mostafazadeh1}
was selected. The positive and negative frequency components are separately
biorthogonal, but they do not propagate causally. In this Section we will show
that sums of positive and negative frequency position eigenvectors are
covariant position eigenvectors that do propagate causally.

Nonlocality due to the similarity transformation to the NW basis is completely
eliminated in the biorthogonal formalism. The Philips states \cite{Philips}
used by Marolf and Rovelli to model Lorentz invariant detectors \cite{Rovelli}
combined with their duals are biorthogonal. While the NW position operator is
Hermitian and the NW basis is equivalent to the biorthogonal basis in the
sense that scalar products are preserved, its eigenvectors are nonlocal in
configuration space due to the Fourier space factor $\omega_{\mathbf{k}}%
^{1/2}.$ In addition, the NW representation requires redefinition of the field
operators, while the biorthogonal representation is based on the canonical
field operators so it can be compared directly to the standard localization
scheme \cite{Halvorson}.

There is a source of nonlocality that is intrinsic to all one-particle
positive frequency states. According to the Hegerfeldt theorem
\cite{Hegerfeldt}, positive frequency states initially localized in a finite
region will spread instantaneously to fill all of space. Physically this is
because localization of positive frequency waves is due to destructive
interference between intrinsically nonlocal counterpropagating waves
\cite{Prigogine}. If negative frequency states are excluded, there are no
localized states that evolve causally \cite{Redhead,Halvorson}. According to
the RS theorem the global vacuum is cyclic for every local algebra. As a
consequence, every local event has a nonzero probability of occurring in the
vacuum and local creation, annihilation and number operators do not exist. For
example, $\widehat{\phi}^{-}\left(  x\right)  \left\vert 0\right\rangle $ is
intrinsically nonlocal, so $\widehat{\phi}^{-}\left(  x\right)  $ is not a
local operator. As discussed in Halvorson's critique of NW localization
\cite{FlemingAdo,Halvorson}, the use of positive frequency orthogonal (or
biorthogonal) bases does not eliminate these consequences of the RS theorem,
it merely masks them for an instant. True localization requires a sum over
positive and negative frequencies \cite{Leon}.

The derivations in Sections 3 and 4 can be applied to a first or a second
quantized theory, but we will start by discussing of the simpler case of first
quantized fields which for photons are just "classical" solutions to Maxwell's
wave equation. For this case biorthogonality of the basis vectors can be
verified directly without reference to the canonical commutation relations,
annihilation and creation operators or the vacuum. Neutral KG particles and
photons are described by real functions of $x$ that propagate causally. What
is new here is that the configuration space bases (\ref{phi_epsilon}) and
(\ref{A_loc}) lead to the particle densities (\ref{p_KG}) and
(\ref{probability}) where $\left\langle \psi|\widetilde{\psi}\right\rangle $
is the total number of particles.

Annihilation and creation operators are intrinsic to any second quantized
theory, and we have seen that these operators are nonlocal. To salvage the
concept of local measurements Knight defined strict localization as
indistinguishability from the vacuum in spacelike separated regions so that
$\left\langle \psi\left\vert \widehat{Q}\right\vert \psi\right\rangle
=\left\langle 0\left\vert \widehat{Q}\right\vert 0\right\rangle $
\cite{Knight}. Here $\widehat{Q}$ is some function of the field operators
$\widehat{\phi}\left(  x_{i}\right)  $ and $\widehat{\pi}\left(  x_{i}\right)
$ and these operators commute at spacelike separated $x_{i}$. If $g\left(
x\right)  $ is a localized solution to the KG equation and%
\begin{equation}
\widehat{R}=\int d\mathbf{x}\left[  g\left(  x\right)  \partial_{ct}%
\widehat{\phi}\left(  x\right)  -\widehat{\phi}\left(  x\right)  \partial
_{ct}g\left(  x\right)  \right]  ,
\end{equation}
then $\left\vert \psi\right\rangle =\exp\left(  \mathrm{i}\widehat{R}\right)
\left\vert 0\right\rangle $ is strictly localized. This is an interesting
state, since it is the coherent field operator radiated by a classical current
distribution \cite{ScullyZubairy}. However, all terms in $\left\vert
\Psi\right\rangle $ are positive frequency and no such state containing a
finite number of particles is strictly localized.

Energy must certainly be bounded from below \cite{Malament}, but we propose
that positive and negative frequencies can be included if both absorbed and
emitted particles are counted on a spacelike hyperplane. The motivation for
this interpretation was the photon-matter interaction, $\widehat{\mathbf{j}%
}\left(  x\right)  \cdot\widehat{\mathbf{A}}\left(  x\right)  $. Particle
position is defined here as the spacetime coordinate $x$ appearing in the
vector potential $\widehat{\mathbf{A}}\left(  x\right)  $ and its derivatives,
or in the field operators $\widehat{\phi}\left(  x\right)  $ and
$\widehat{\pi}\left(  x\right)  $. In QFT these fields describe absorption and
emission of photons by an atom or of neutral KG particles by an Unruh-Davies
monopole detector \cite{BirrellDavies}. The positive frequency parts of
$\widehat{\mathbf{A}}\left(  x\right)  $ and $\widehat{\phi}\left(  x\right)
$ describes absorption of positive energy particles, while their negative
frequency parts describe emission of positive energy particles.

We will show next that this interpretation is consistent with microcausality
which is the requirement that observables be described by field operators that
commute for spacelike separated events: The relationship of microcausality to
inclusion of negative frequencies in the scalar product (\ref{x_sp}) can be
seen by substituting (\ref{phi_epsilon}) to (\ref{negative}) in the vacuum
expectation value of (\ref{x_commutation}) to give
\begin{align}
\left\langle \phi\left(  x\right)  |\pi\left(  y\right)  \right\rangle  &
=\left\langle 0\left\vert \widehat{\phi}^{+}\left(  x\right)  \widehat{\pi
}^{-}\left(  y\right)  -\widehat{\pi}^{+}\left(  y\right)  \widehat{\phi}%
^{-}\left(  x\right)  \right\vert 0\right\rangle \nonumber\\
&  =\left\langle \phi^{+}\left(  x\right)  |\pi^{+}\left(  y\right)
\right\rangle {+}\left\langle \pi^{+}\left(  y\right)  |\phi^{+}\left(
x\right)  \right\rangle \nonumber\\
&  =\left\langle \phi^{+}\left(  x\right)  |\pi^{+}\left(  y\right)
\right\rangle {+}\left\langle \phi^{-}\left(  x\right)  |\pi^{-}\left(
y\right)  \right\rangle . \label{c}%
\end{align}
On the $t_{x}=t_{y}$ hyperplane events $x^{\mu}=\left(  ct_{x},\mathbf{x}%
\right)  $ and $y^{\mu}=\left(  ct_{y},\mathbf{y}\right)  $ appear
simultaneous but an inertial observer with velocity $c\beta$ will see these
events as time ordered. Since the Fourier space integrand of $\left\langle
\phi^{+}\left(  x\right)  |\pi^{+}\left(  y\right)  \right\rangle $ is
proportional to $e^{-\mathrm{i}\omega_{\mathbf{k}}\left(  t_{x}-t_{y}\right)
}$ while that of $\left\langle \pi^{-}\left(  y\right)  |\phi^{-}\left(
x\right)  \right\rangle $ will be seen as proportional to $e^{\mathrm{i}%
\omega_{\mathbf{k}}\left(  t_{x}-t_{y}\right)  },$ if $t_{x}>t_{y}$ the first
term is positive frequency $\left(  \epsilon=+\right)  $ while the second is
negative frequency $\left(  \epsilon=-\right)  $. This assignment is not
unique, since an observer with velocity $-c\beta$ will see the opposite time
order. Thus the manifestly covariant scalar product (\ref{c}) is a sum over
absorbed and emitted particles as in \cite{HawtonPhysicaScripta}.

The basis states (\ref{phi_epsilon}) and the scalar product
(\ref{invariant_sp}) are directly related to the QFT Green functions as
defined in \cite{HalliwellOrtiz}. These propagators satisfy the KG equation
(\ref{KGeq}) with a $\delta^{4}\left(  x-y\right)  $ source and are of the
form%
\begin{equation}
\mathcal{G}\left(  x-y\right)  =\frac{1}{\left(  2\pi\right)  ^{3}}\int
d^{4}kf\left(  k^{0}\right)  \delta\left(  k^{2}-m^{2}\right)  e^{-\mathrm{i}%
k\left(  x-y\right)  }.
\end{equation}
The choice $f\left(  k^{0}\right)  =\Theta\left(  k^{0}\right)  $ gives the
positive and negative frequency Wightman functions $G^{+}\left(  x-y\right)  $
and $G^{-}\left(  x-y\right)  =G^{+}\left(  y-x\right)  ,$ $f\left(
k^{0}\right)  =-\mathrm{i}\epsilon\left(  k^{0}\right)  $ gives the causal or
commutator function $G\left(  x-y\right)  $ for which $\mathrm{i}G\left(
x-y\right)  =G^{+}\left(  x-y\right)  -G^{-}\left(  x-y\right)  $ and
$\partial_{ct_{x}}G\left(  x-y\right)  |_{t_{x}=t_{y}}=-\delta\left(
\mathbf{x}-\mathbf{y}\right)  ,$ and $f\left(  k^{0}\right)  =1$ gives the
Hadamard/Schwinger Green function for which $G^{\left(  1\right)  }\left(
x-y\right)  =G^{+}\left(  x-y\right)  +G^{-}\left(  x-y\right)  .$ These
propagators can be related to the functions defined in Sections 2 and 3 by
identifying
\begin{equation}
\psi_{y}^{\epsilon}\left(  x\right)  \equiv\left\langle \pi^{\epsilon}\left(
x\right)  |\phi^{\epsilon}\left(  y\right)  \right\rangle =\mathrm{i}%
\epsilon\frac{\hbar}{2}\partial_{ct_{x}}G^{\epsilon}\left(  x-y\right)
\end{equation}
as the probability amplitude for absorption at $x$ of a particle emitted by a
source at $y$ $\left(  \epsilon=+\right)  ,$ or absorption at $y$ of a
particle emitted at $x$ $\left(  \epsilon=-\right)  $. The $\epsilon
$-frequency potential-like field at $x$ due to a localized source at $y$ is
$\phi_{y}^{\epsilon}\left(  x\right)  \equiv\left\langle \phi^{\epsilon
}\left(  x\right)  |\phi^{\epsilon}\left(  y\right)  \right\rangle $ so
$G^{\left(  1\right)  }\left(  x-y\right)  =\phi_{y}^{+}\left(  x\right)
+\phi_{y}^{-}\left(  x\right)  $ and \textrm{i}$G\left(  x-y\right)  =\phi
_{y}^{+}\left(  x\right)  -\phi_{y}^{-}\left(  x\right)  $ are the field and
its conjugate defined in Section 2. The position eigenvectors evolve in time
like derivatives of the corresponding QFT propagators. With $t\equiv
t_{x}-t_{y}$ and $r\equiv\left\vert \mathbf{x}-\mathbf{y}\right\vert ,$
\begin{align}
\psi_{y}^{\epsilon}\left(  x\right)   &  =\int\frac{d\mathbf{k}}{\left(
2\pi\right)  ^{3}2}e^{-\mathrm{i}\epsilon\left[  \omega_{\mathbf{k}}\left(
t_{x}-t_{y}\right)  -\mathbf{k\cdot}\left(  \mathbf{x}-\mathbf{y}\right)
\right]  }\label{Psi_epsilon}\\
&  \underset{m=0}{=}\frac{-1}{8\pi^{2}r}\frac{\partial}{\partial r}%
\sum_{\gamma=\pm}\left[  \pi\delta\left(  r-\gamma\epsilon ct\right)  +i\gamma
P\left(  \frac{1}{r-\gamma\epsilon ct}\right)  \right]  \text{ }\nonumber
\end{align}
where $\mathbf{k}$ was replaced with $-\mathbf{k}$ in the $\epsilon=-$ term.
At $t=0$ the delta functions add while the principal values terms cancel so
$\psi_{y}^{\epsilon}\left(  x\right)  $ is localized. However, for $t_{x}\neq
t_{y},$ $\psi_{y}^{\epsilon}\left(  x\right)  $ has a nonlocal imaginary part
that leads to instantaneous spreading. Any state vector can be expanded in the
$\left\vert \phi^{\epsilon}\left(  x\right)  \right\rangle $ basis with wave
function components $\psi^{\epsilon}\left(  x\right)  $ given by (\ref{wf}) or
in the $\left\{  \left\vert \phi_{c}\left(  x\right)  \right\rangle
,\left\vert \phi\left(  x\right)  \right\rangle \right\}  /\sqrt{2}$ basis
with wave function components $\psi_{c}\left(  x\right)  =\left[  \psi
^{+}\left(  x\right)  +\psi^{-}\left(  x\right)  \right]  /\sqrt{2}$ and
$\psi\left(  x\right)  =\left[  \psi^{+}\left(  x\right)  -\psi^{-}\left(
x\right)  \right]  /\sqrt{2}$. For the state (\ref{Psi_epsilon}) only
$\psi_{yc}\left(  x\right)  =\frac{\hbar}{2}\partial_{ct_{x}}G\left(
x-y\right)  $ which equals $-\frac{\hbar}{2}\delta\left(  \mathbf{x-y}\right)
$ at $t_{x}=t_{y}$ is a position eigenvector. These basis states satisfy the
boundary conditions $\phi_{yc}\left(  x\right)  =0$ and $\partial_{ct_{x}}%
\phi_{yc}\left(  x\right)  =\delta\left(  \mathbf{x-y}\right)  $ on the
$t_{x}$ hyperplane and provide a basis for the real fields that are consistent
with microcausality. For $t_{x}>t_{y}$ $\phi_{yc}\left(  x\right)  $
propagates outward on the spherical shell $r=ct,$ while for $t_{x}<t_{y}$ it
propagates inward on $r=-ct$. The positive and negative frequency Wightman
function are covariant in the senses that they can evaluated on an arbitrary
spacelike hyperplane \cite{PS}, but this does not take into account the
observer dependence of time ordering that is reflected in the sign of the
exponent in (\ref{Psi_epsilon}). A sum over positive and negative frequencies
is required for causal time evolution and observer independence of the wave
function with time order taken into account. According to (\ref{c}) emission
at $y$ with absorption at $x$ and emission at $x$ with absorption at $y$ are
correlated in the $\epsilon=\pm$ basis.

Time development is described by the operator $\widehat{U}\left(  t\right)
=\exp\left(  -\mathrm{i}\widehat{\omega}t\right)  $ where $\widehat{\omega
}_{\epsilon\epsilon^{\prime}}=\epsilon c\widehat{D}^{1/2}\delta_{\epsilon
\epsilon^{\prime}}$ are the elements of the $2\times2$ matrix that generates
infinitesimal displacements in time \cite{Mostafazadeh1}. The symbol
$\widehat{\omega}$ is used here to emphasize that it is a frequency, not an
energy, operator. Its spectrum, $\epsilon c\left\vert \mathbf{k}\right\vert ,$
is not bounded below. It is this property that allows the states
(\ref{Psi_epsilon}) to escape the nonexistence theorem proved in
\cite{Malament} and consequences of the antilocal operator $\widehat{D}%
^{-1/2}$ discussed by Halvorson \cite{Halvorson}. The spectrum of the energy
operator, $\widehat{H}_{\epsilon\epsilon^{\prime}}=\hbar c\widehat{D}%
^{1/2}\delta_{\epsilon\epsilon^{\prime}},$ is bounded below. Hegerfeldt
assumed positive energies which we identify here with positive frequencies, so
the consequences of his theorem are avoided.

The field due to a distributed source is $\phi\left(  x\right)  =\int
dyG\left(  x-y\right)  j\left(  y\right)  $. The causal behavior of space-like
separated particle devices described by Eq. (25) of \cite{MM} is of this form.
For photons interacting with charged matter $J$ and $\phi$ should be replaced
with the four-vectors $J^{\mu}$ and $A^{\mu}.$ If restricted to positive
frequencies, the wave function $\mathbf{\psi}_{\lambda}\left(  x\right)
=\left\langle \mathbf{E}_{\lambda}^{+}\left(  x\right)  |\psi\left(  t\right)
\right\rangle $ emitted by an atom does not evolve causally due to the factor
$\Theta\left(  k^{0}\right)  $ \cite{DebierreDurt}. However the field due to
the current $\mathbf{j}_{e}\left(  y\right)  =\left\langle \psi\left\vert
\widehat{\mathbf{j}}_{e}\left(  y\right)  \right\vert \psi\right\rangle $ with
$\widehat{\mathbf{j}}_{e}=\frac{e}{m_{e}}\widehat{\mathbf{p}}_{e}\delta\left(
\mathbf{y}\right)  $ and $\left\vert \psi\left(  t\right)  \right\rangle $
given by (\ref{eq:IPState}) can be written as a integral over the causal
propagator $G\left(  x-y\right)  $ which contains no factor $\Theta\left(
k^{0}\right)  .$ The atomic source current at time $t_{y} $ due to a dipole
localized at $\mathbf{y}=0$ is $\mathbf{j}_{e}\left(  y\right)  =\mathbf{j}%
_{e}\delta\left(  \mathbf{y}\right)  \Theta\left(  t_{y}\right)  \exp\left(
-\mathrm{i}\Omega_{0}t_{y}\right)  $ where $\Omega_{0}=\omega_{0}+\omega
_{LS}-\mathrm{i}\Gamma/2,$ $\Gamma$ is the decay rate and $\omega_{LS}$ is the
partial Lamb shift \cite{DebierreDurt}. The wave function emitted by this
current source observed at time $t_{x}$ is then proportional to $\mathbf{j}%
_{e}\Theta\left(  t_{x}-\left\vert \mathbf{x}\right\vert /c\right)
\exp\left[  -\mathrm{i}\Omega_{0}\left(  t_{x}-\left\vert \mathbf{x}%
\right\vert /c\right)  \right]  $ so it propagates causally
\cite{DebierreDurt}. Negative frequencies are allowed since there are local
vacuum fluctuations correlated across spacelike separated regions even in the
global vacuum. In the covariant localized basis absorption and emission are
equally likely, but the wave function of an atom known to be initially excited
in the global vacuum is the sum of a large positive frequency emission term
and a very small negative frequency absorption term \cite{DebierreDurt} that
here is attributed to local vacuum fluctuations. While absorption and emission
are equally likely in the basis vectors, knowledge that the atom was initially
in its excited state in the global vacuum leads a device dominated by emission.

Localization and causality is more complicated in the case of photons
\cite{Debierre}. The completeness relation (\ref{complete}) can be used to
expand the first quantized vector potential in the coordinate space basis as
in \cite{BabaeiMostafazadeh} or to expand the state vector in the
configuration space basis as in our Sections 4 and 5. In the former case the
probability amplitude to find the photon at $x$ is proportional to the scalar
$\sum_{i}\left\langle E_{\lambda i}^{\epsilon}\left(  x\right)  |A_{i}%
\right\rangle .$ If $A_{i}$ is the position eigenvector $A_{\lambda
i}^{\epsilon}\left(  y\right)  ,$ this is\ given by (\ref{orthogonal}) so it
is localized at $y$. This is consistent with the wave function defined in
\cite{BabaeiMostafazadeh}. If the state vector is expanded, the wave function
is a transverse vector and a covariant description requires the Lorenz gauge.
In this gauge the causal photon Green function is $g^{\mu\nu}G\left(
x-y\right)  $ where $g^{\mu\nu}=\mathrm{diag}\left(  1,-1,-1,-1\right)  $ and
the source for $A^{\mu}\left(  x\right)  $ is a conserved four-current
\cite{MandlShaw,CohenQED1}. All components of $A^{\mu}\left(  x\right)  $
propagate causally on the light cone. In a source free region where the photon
is on its mass shell the longitudinal part of the electric field
$\mathbf{E}\left(  x\right)  =-\partial_{t}\mathbf{A}\left(  x\right)
-\mathbf{\nabla}\phi\left(  x\right)  $ is zero due to the Lorenz gauge
condition so $\mathbf{E}\left(  x\right)  $ and $\mathbf{B}\left(  x\right)
=\mathbf{\nabla}\times\mathbf{A}\left(  x\right)  $ are transverse.
Propagation is causal and the photon position eigenvectors are localized. As
discussed in the Introduction, the photon position eigenvectors have defined
total angular momentum along their axis of cylindrical symmetry. All of the
commutation relations of the Poincar\'{e} group are satisfied as discussed in
\cite{HawtonPosOp,Hawton07} where the relevant case for the biorthogonal basis
is $\alpha=0$. The spacetime location of a position eigenvector is defined as
the argument $x$ in $\widehat{A}^{\mu}\left(  x\right)  $, the internal
angular momentum of a position eigenvector includes the orbital angular
momentum relative to $\mathbf{x}$ and the external angular momentum operator
is $\widehat{\mathbf{x}}\times\widehat{\mathbf{p}}.$

The position eigenvectors describe a particle at position $\mathbf{x}$, but it
is uncertain whether it was annihilated or created there. Only $\mathbf{k}
$-space is a true Fock space so particles must be created and annihilated
globally. The KG and photon position eigenvectors (including the negative
frequency ones) are given in terms of this Fock basis in (21-22) and (48-51)
respectively. This is consistent with the Reeh-Schlieder conclusion that there
are no local creation or annihilation operators. For a detector array arranged
to capture all photons of interest, locally there is only probability density
for the presence of a photon at $\mathbf{x}$ on some hyperplane. The
experimenter must examine the whole array to count photons. When applied
globally, the number operator \cite{NumDens95}%
\begin{align}
\widehat{N}_{\lambda}  &  =\mathrm{i}\frac{\epsilon_{0}}{\hbar}\int%
\mathrm{d}\mathbf{xE}_{\lambda}^{-}\left(  \mathbf{x},t\right)  \cdot
\mathbf{A}_{\lambda}^{+}\left(  \mathbf{x},t\right)  +H.c.\label{N}\\
&  =\int\frac{\mathrm{d}\mathbf{k}}{\left(  2\pi\right)  ^{3}2\omega
_{\mathbf{k}}}\widehat{a}_{\lambda}^{\dagger}\left(  \mathbf{k}\right)
\widehat{a}_{\lambda}\left(  \mathbf{k}\right) \nonumber
\end{align}
counts photons with helicity $\lambda$ at time $t.$

\section{Conclusion}

The formalism of biorthogonal systems can be, as we saw, called in action in
relativistic quantum mechanics. It is particularly well-matched to the
relativistic scalar product. In the biorthogonal formalism, both the
Wigner-Bargmann quantum field operator (for photons, the vector potential) and
its canonically conjugate momentum (for photons, the electric field) are put
on an equal footing, and they generate respectively the direct and the dual
basis of position eigenvectors of two different position operators, which are
the Hermitian conjugate of each other. Our formalism further clarifies the
meaning of the free parameter $\alpha$ \cite{HawtonBaylis,Hawton07,Debierre}
in the photon position operator.

The probability density (\ref{probability}) suggests a resolution of the
apparent dichotomy between photon number counting and the sensitivity of a
detector to energy density. The wave function $\left\langle \mathbf{E}%
_{\lambda}^{+}\left(  x\right)  |\psi\left(  t\right)  \right\rangle $
together with the state vector (\ref{eq:IPState}) describes creation of a
photon in the time interval $0\leq t^{\prime}\leq$ $t$ followed by its
detection at time $t.$ Since it is created in the $\left\vert \mathbf{A}%
_{\lambda}^{\epsilon}\left(  x\right)  \right\rangle $ basis and observed in
the dual $\left\vert \mathbf{E}_{\lambda}^{\epsilon}\left(  x\right)
\right\rangle $ basis, that wave function is proportional to a probability
amplitude. The probability density for a transition from $\left\vert
\psi_{\perp}\left(  t\right)  \right\rangle $ to the position eigenvector at
$\mathbf{x}$, given by $\frac{2\epsilon_{0}}{\hbar}\left\vert \left\langle
\mathbf{E}_{\lambda}^{+}\left(  x\right)  |\psi\left(  t\right)  \right\rangle
\right\vert ^{2},$ is of the Glauber form \cite{Glauber}. However, in contrast
to theories of photodetection based on energy density, we have proposed,
through the position amplitude $\left\langle \mathbf{E}_{\lambda}^{+}\left(
x\right)  |\psi\left(  t\right)  \right\rangle $ in the dual basis, a true
position measurement that describes an array of ideal photon counting detectors.

For a state vector that is an arbitrary linear combination of positive and
negative frequency terms the probability density for a transition to the
position eigenvector at $x${\ is positive definite. This probability density
describes a particle at spatial location }$x${\ independent of whether it was
absorbed or emitted. Thus (\ref{probability}) can be interpreted as
probability density even if the wave function (\ref{wave_function}) is real as
in classical electromagnetism. }This application of biorthogonal QM is based
on an invariant positive definite scalar product so transition probabilities
are invariant and positive definite, the position operator is covariant, and
there is no NW $\omega_{\mathbf{k}}^{\pm1/2}$ nonlocality in the wave
function. In any theory that combines relativity with QM the position
coordinate is problematic. In QM a position measurement should be associated
with a Hermitian position operator, but the only orthogonal position
eigenvectors available in the published literature are nonlocal and
noncovariant \cite{NewtonWigner}. The NW construction can be extended to
photons [34] but then coupling of photon number amplitude to current density
is nonlocal \cite{Cook}. In the biorthogonal formalism number density is
derived from the canonical fields and it becomes clear that the relationships
amongst number, energy and current density are all local. If the NW basis is
restricted to the calculation of scalar products and expectation values there
is no disagreement between the conventional treatment of particle position
based on the NW position operator \cite{Schweber} and biorthogonal QM.
However, one should not fall into the trap of believing that NW nonlocality is
physically real and hence observable.


\begin{thebibliography}{99}
\bibitem {quasi}Quasi-Hermitian operators are pseudo-Hermitian operators
restricted to a positive definite norm as required for the probability
interpretation of QM \cite{GeyerScholtz}.
\end{thebibliography}
\end{document}